# Quantitative phase imaging through an ultra-thin lensless fiber endoscope

**Authors**

Jiawei Sun,[1,2]* Jiachen Wu,[1,3] Song Wu,[4] Ruchi Goswami,[5] Salvatore Girardo,[5] Liangcai Cao,[3] Jochen Guck,[5,6] Nektarios Koukourakis,[1,2]* and Juergen W. Czarske[1,2,6,7]*

**Affiliations**

1. Laboratory of Measurement and Sensor System Technique (MST), TU Dresden, Helmholtzstrasse 18, 01069 Dresden, Germany.
2. Competence Center for Biomedical Computational Laser Systems (BIOLAS), TU Dresden, Dresden, Germany.
3. State Key Laboratory of Precision Measurement Technology and Instruments, Department of Precision Instruments, Tsinghua University, 100084 Beijing, China.
4. Institute for Integrative Nanosciences, IFW Dresden, Helmholtzstraße 20, 01069 Dresden, Germany
5. Max Planck Institute for the Science of Light & Max-Planck-Zentrum für Physik und Medizin, 91058 Erlangen, Germany.
6. Cluster of Excellence Physics of Life, TU Dresden, Dresden, Germany.
7. Institute of Applied Physics, TU Dresden, Dresden, Germany.
* [jiawei.sun@tu-dresden.de](jiawei.sun@tu-dresden.de) (J.S.); [nektarios.koukourakis@tu-dresden.de](nektarios.koukourakis@tu-dresden.de) (N.K.); [juergen.czarske@tu-dresden.de](juergen.czarske@tu-dresden.de) (J.C.)

**Abstract**

Quantitative phase imaging (QPI) is a label-free technique providing both morphology and quantitative biophysical information in biomedicine. However, applying such a powerful technique to *in vivo* pathological diagnosis remains challenging. Multi-core fiber bundles (MCFs) enable ultra-thin probes for *in vivo* imaging, but current MCF imaging techniques are limited to amplitude imaging modalities. We demonstrate a computational lensless microendoscope that uses an ultra-thin bare MCF to perform quantitative phase imaging with microscale lateral resolution and nanoscale axial sensitivity of the optical path length. The incident complex light field at the measurement side is precisely reconstructed from the far-field speckle pattern at the detection side, enabling digital refocusing in a multi-layer sample without any mechanical movement. The accuracy of the quantitative phase reconstruction is validated by imaging the phase target and hydrogel beads through the MCF. With the proposed imaging modality, three-dimensional imaging of human cancer cells is achieved through the ultra-thin fiber endoscope, promising widespread clinical applications.

## Introduction

QPI is an effective and label-free method for cell and tissue imaging in biomedicine [1]. 3D images of transparent samples can be reconstructed with QPI in a non-invasive manner [2–10], enabling nanoscale sensitivity to morphology and dynamics. Meanwhile, quantitative biophysical parameters such as refractive index [11,12], dry mass [13,14], matter density [15], and skewness [16] can be extracted from the quantitative phase shift, providing both morphological and quantitative biophysical information for digital pathology [17]. Recent research combining QPI with deep learning has been used for virtual staining [18,19] and dynamic blood examination [20,21], which was reported as a high throughput approach to detecting the SARS-CoV-2 virus [22]. On the other hand, current QPI methods are mostly based on bulky and expensive microscope platforms with limited working distance and penetration depth, which means invasive sampling or sectioning of diseased tissues or organs are required for pathological diagnosis [23,24]. Such invasive approaches limit the *in vivo* application of QPI in clinical diagnosis, especially in the early diagnosis of cancer and tumors.

In clinical diagnosis, endoscopes with diameters of a few millimeters are commonly used for *in vivo* imaging. MCF is an ultra-thin fiber bundle of a few hundred micrometers consisting of thousands of single-mode fiber cores (see Fig. 1a, b), and recent advances in MCF-based computational imaging demonstrate the great potential of fiber bundles to be the next generation microendoscopes with minimal invasiveness [25–27]. However, the phase information of the sample is lost due to the incoherent illumination. Despite computational methods that have been proposed to recover the 3D information of samples [28–30], precise QPI via MCF with nanoscale sensitivity is still challenging. Coherent imaging is achieved via multi-mode fibers with transmission matrix measurement [31,32] or wavefront shaping [33–38], and similar approaches are also applied to MCF-based coherent imaging [39–45]. In practice, bulky and expensive optical systems with spatial light modulators and complicated calibration processes are still required, and the scanning-based imaging technique can be slow, inducing many limitations for clinical applications. Furthermore, an endoscope with nanoscale sensitivity of the optical path length is not yet reported, therefore, a simple and cost-effective 3D microendoscope with nanoscale sensitivity is highly demanded.

In this research, we found that the MCF can directly work as a phase encoder without a coded aperture [29] at the measurement side, encoding the incident complex light field to a speckle pattern in the far-field at the detection side. We propose a novel computational approach named the far-field amplitude-only speckle transfer (FAST) method to decode the incident light field from the far-field speckles. Unlike conventional fiber facet imaging methods, where imaging resolution is limited by the core-to-core spacing (Fig. 1c), our approach enables 3D QPI reconstruction with nanoscale axial sensitivity of optical path length and lateral resolution up to 1μm in the ideal case via direct recovery of the incident complex light field (Fig. 1d). We demonstrate a computational quantitative phase microendoscope (QPE) providing both morphological and quantitative biophysical information with a simple optical system (Fig. 1e), paving the path for *in vivo* clinical applications of the fiber bundles.

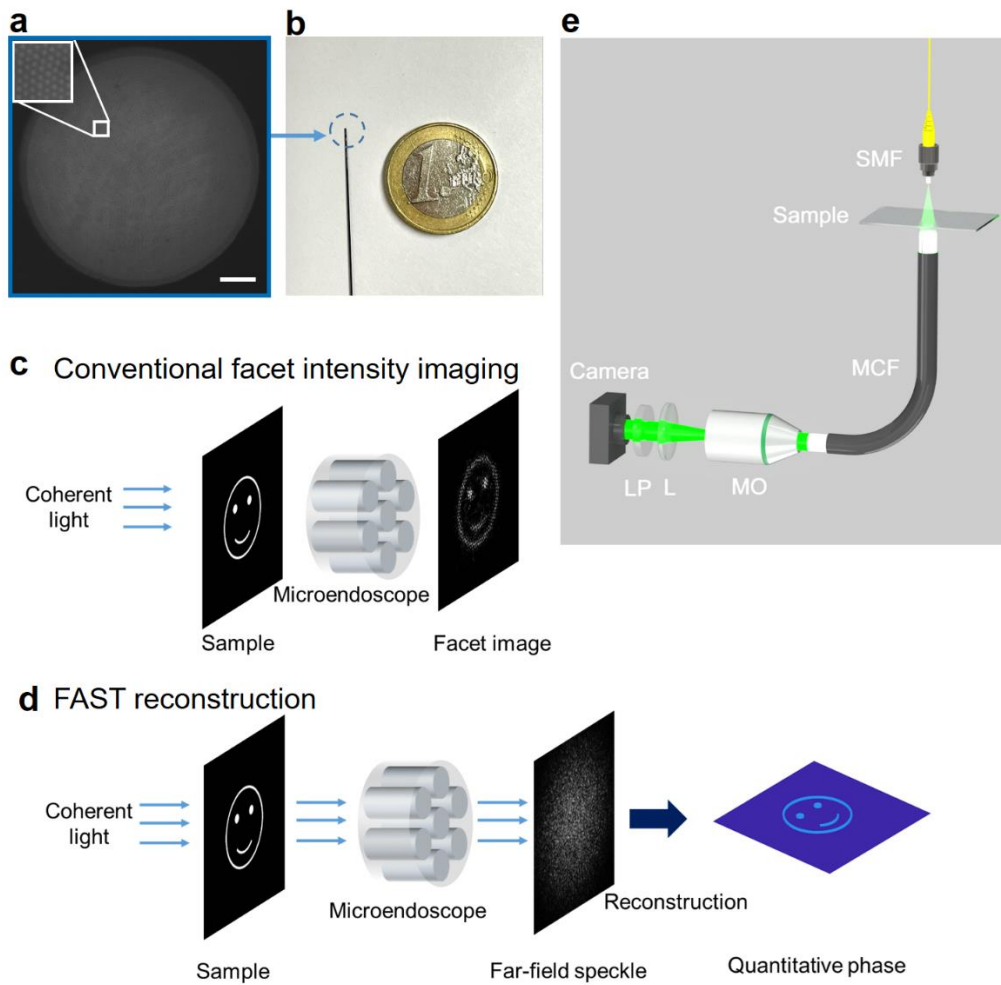

**Fig. 1. Lensless quantitative phase microendoscope: setup and concept.** (**a**) Microscopic image of the tip facet of the 10,000 core fiber bundle with a diameter of 350μm. Scale bar 50μm. (**b**) Photo of the fiber bundle and a one euro coin for scale. (**c**) Conventional lensless microendoscopic imaging can only get the pixelated intensity information of the specimen, and the sample has to be very close to the fiber facet. (**d**) Quantitative phase and high-resolution amplitude images of the specimen can be reconstructed from the far-field speckle image. The sample can be placed far from the facet due to the digital focusing capability. (**e**) Experimental setup; SMF, single-mode fiber; MO, microscope objective; L, achromatic lens; LP, linear polarizer.

## Results
### Image reconstruction through the fiber bundle

In an MCF, the optical path length (OPL) varies for light traveling in different fiber cores, which results in a random phase distribution at the detection side for a plane wave illumination at the measurement side. The intrinsic OPL difference is stable when the fiber bundle is static in the measurement process. The phase shift induced by the sample can thus be reconstructed from intensity-only far-field speckles at the detection side.

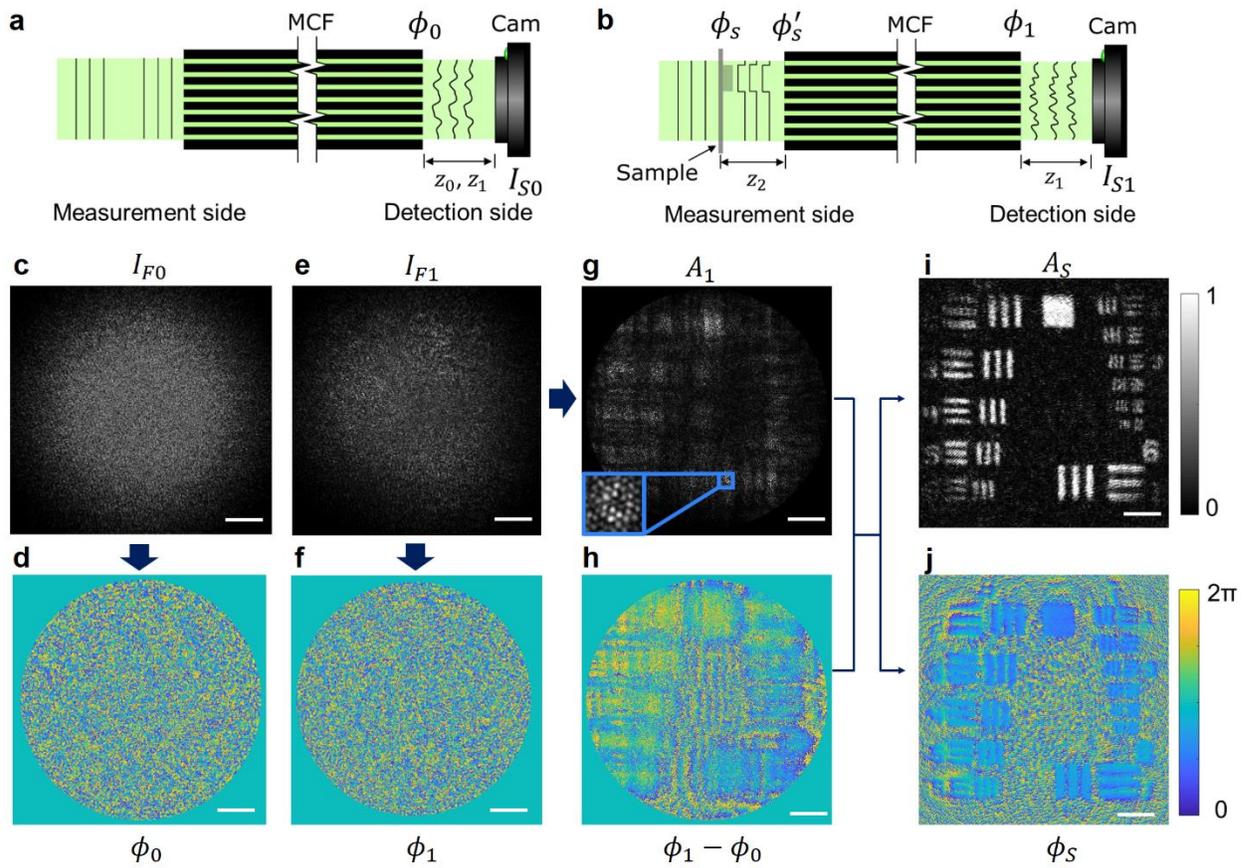

**Fig. 2. Working principle of the lensless quantitative phase microendoscope.** (**a**) The phase of a coherent light source is distorted by the MCF at the detection side. (**b**) A USAF resolution test target is located 1.6mm away from the facet. The far-field speckle image is captured on the camera. (**c**) Reference speckle image captured on the detection side with a point light source illumination on the measurement side. (**d**) Reconstructed reference phase image on the fiber facet. (**e**) Far-field speckle image of the test target. (**f**) Phase and (**g**) amplitude reconstruction on the fiber facet, which contains the light field information of the test target. (**h**) Retrieved phase image of the diffracted test target. Reconstructed (**i**) amplitude and (**j**) phase image of the 6th and 7th group elements of the test target on the focal plane. Scale bars 50μm.

The imaging principle and reconstruction process of the lensless quantitative phase microendoscope is demonstrated in Fig. 2. Initially, the MCF is illuminated by a collimated laser beam or a point light source for a reference measurement of the intrinsic OPL difference of fiber cores. Two far-field speckle patterns, which are $z_0$ and $z_1$ away from the fiber facet at the detection side, are magnified and projected on the camera at the detection side (Fig. 2a). The intrinsic phase shift of the MCF (Fig. 2d) induced by the OPL difference is reconstructed from the far-field speckles (Fig. 2c) with the FAST algorithm (see supplementary materials).

A negative resolution test target, where only the pattern is transparent, is put 1.6mm ($z_2$) away from the facet as a sample at the measurement side (Fig. 2b). The speckle pattern $z_1$ away from the facet, which is the system response of the sample, is captured on the camera at the detection side (Fig. 2e). The phase $\phi_1(x, y)$ and amplitude $A_1(x, y)$ information on the facet is reconstructed from the intensity-only far-field speckle. The phase of the sample is encoded at the detection side as shown in Fig. 2f due to the fiber core OPL difference. The original phase incident on the fiber bundle at the measurement side $\phi'_s(x, y)$, which contains the phase information of the test target, can be decoded by the measured intrinsic phase shift $\phi_0(x, y)$ as shown in Fig. 2h.

$$\phi'_s = \phi_1 - \phi_0 \tag{1}$$

On the other hand, the original amplitude information of the incident light field is maintained at the detection side as shown in Fig. 2g. Therefore, the incident light field on the measurement facet can be expressed as a complex field $E'_s(x, y)$.

$$E'_s = A_1 \cdot \exp(i\phi'_s) \tag{2}$$

Hence, the incident light field is back-propagated numerically to the sample plane with the angular spectrum method [46]. The digital-focused amplitude and phase image of the test target is calculated from the propagated complex field. The 6 and 7 group elements of the test chart are resolved in both amplitude (Fig. 2i) and phase (Fig. 2j) reconstruction through the MCF. It can be noticed that the field of view is further extended beyond the size of the fiber facet with the FAST technique.

**Digital refocusing**

Two stacked positive resolution test targets, where the patterns are not transparent, are used to characterize the digital refocusing capability of the proposed microendoscope system. The axial distance between the patterns on the test targets is 1.4mm, and the top layer is 1.26mm away from the facet at the measurement side. On the detection side, the 3D information of the light field is stored in the far-field speckle (Fig. 3a). The complex field of the incident light at the measurement side is reconstructed from the proposed method. Hence, the slices at various axial distances are recovered from the complex field by numerical propagation (Fig. 3b). A reconstructed video demonstrating the digital focusing process is shown in Supplementary video V1.

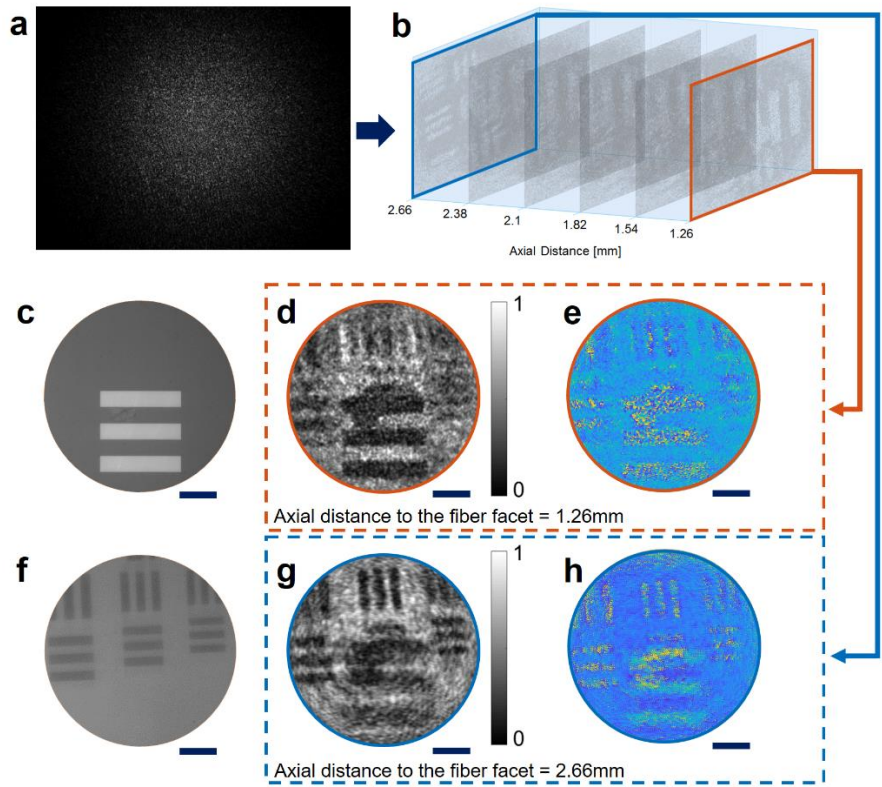

**Fig. 3. Reconstruction of a multi-layer target, which consists of two test targets located at 1.26mm and 2.66mm away from the facet at the measurement side.** (**a**) Speckle pattern of the multi-layer target captured on the detection camera. (**b**) Reconstructed amplitude slides of the sample at different axial distances. (**c**) Image of the top layer from a bulky reflective microscope. (**d-e**) Reconstructed (**d**) amplitude and (**e**) phase image of the top layer, the linewidth is 22.1μm. (**f**) Image of the bottom layer from a bulky reflective microscope. (**g-h**) Reconstructed (**g**) amplitude and (**h**) phase image of the bottom layer, the linewidths are 11.05, 9.84, and 8.77μm respectively. Scale bars 50μm.

Images of the stacked test targets at both layers can be also acquired with a bulky reflective microscope (Fig. 3c, f), but mechanical tuning of the sample position is required. In contrast, images of samples located at multiple axial distances (Fig. 3d, g) can be reconstructed from a speckle image captured at the detection side of the MCF. The phase information can also be recovered at different axial distances as shown in Fig. 3e, h. The lines on the test target are not transparent, which leads to the random phase distribution in the area of lines.

**Glass bead flow reconstruction**

A reconstructed video demonstrating the glass bead flow in a microchannel which is imaged through the microendoscope is demonstrated in Supplementary video V2. The glass bead suspension is pumped into a microchannel constantly by a syringe. The channel is located 1mm away from the fiber facet at the measurement side. The corresponding far-field speckles at the detection side are recorded on the camera at a frame rate of 10 frames per second, and all the frames are processed offline.

## Quantitative phase imaging reconstruction

Phase imaging can provide additional contrast in label-free microscopic imaging, and precise measurement of quantitative phase values can define the refractive index or thickness of biomedical samples. Due to the nanoscale sensitivity of the optical path length, the proposed system has the potential to further measure the height of nanoscale semiconductor structures.

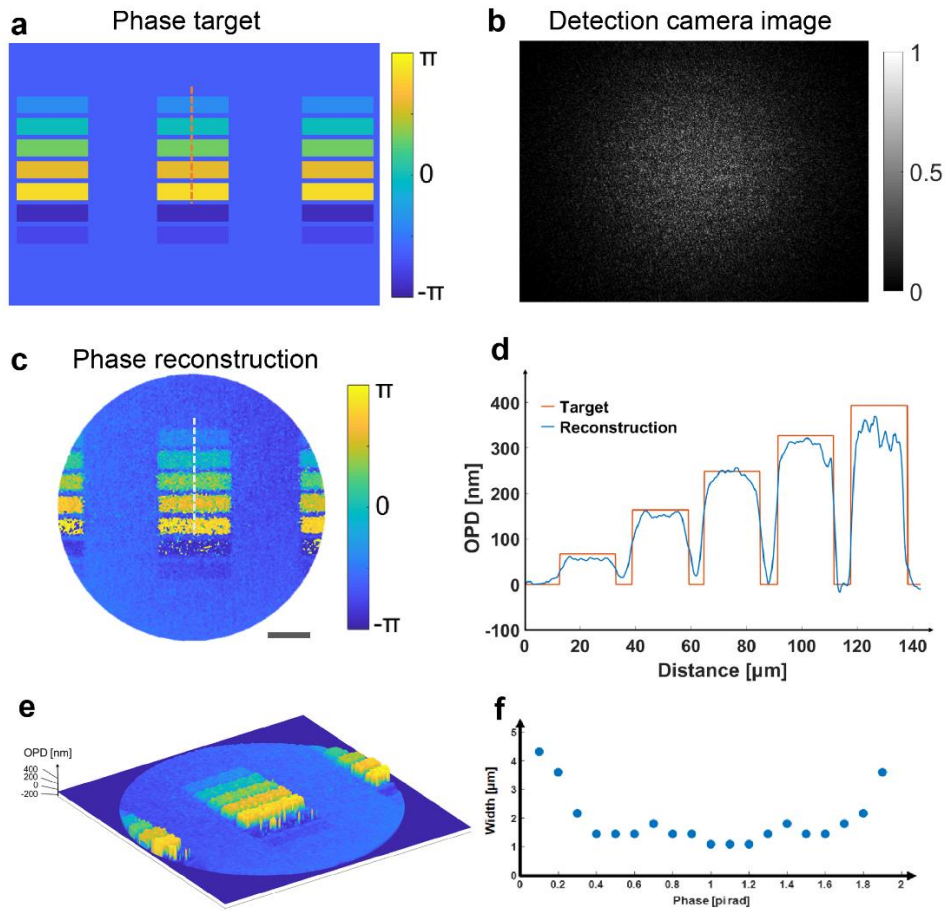

**Fig. 4. Quantitative phase imaging through the microendoscope via far-field speckle reconstruction.** (**a**) The phase target is imaged on the distal side of the microendoscope. (**b**) The far-field speckle image of the phase target is captured on the camera. (**c**) Reconstructed phase image of the target from the speckle image. Scale bar 50μm. (**d**) Quantitative optical path difference profile of the marked areas in (a, c). (**e**) 3D optical path difference map. (**f**) The measured lateral resolution of the system for different phase values. The measurement process is demonstrated in the supplementary materials.

A phase target shown in Fig. 4a is used to characterize the precision of the reconstructed phase from the MCF microendoscope. The phase target is projected on the MCF at the measurement side, and the corresponding system response on the detection camera is shown in Fig. 4b. The quantitative phase image is reconstructed from the speckle image with the FAST method. The phase tilt in the background is corrected numerically and a simulated phase mask with the same phase tilt is subtracted to correct the phase value in the background. The final quantitative phase reconstruction is demonstrated in Fig. 4c. Colors in the phase image represent different phase values, hence, the quantitative phase information is successfully recovered with the FAST reconstruction. A vital optical parameter - optical path difference (OPD), which correlates the refractive index and the thickness of the sample, can be calculated from the quantitative phase shift. A comparison of the calculated OPD between the original phase target and the phase reconstruction through the fiber bundle is demonstrated in Fig. 4d, characterizing the high fidelity of the quantitative phase reconstruction. The data colored in orange indicates the data sampled from the phase target and the blue represents the sampled data from the phase reconstruction. The size of the sample area is 142μm in the longitudinal direction and 2μm in the lateral direction. The calculated 3D OPD map is demonstrated in Fig. 4e and Supplementary video V3.

To further characterize the resolution limit of the system, a program-controlled phase target with tunable phase value and size is implemented. A detailed explanation of the measurement process is demonstrated in supplementary materials. As shown in Fig. 4f, due to the relatively low signal-to-noise ratio (SNR) for the target with an absolute phase shift lower than $0.4\pi$, the lateral resolution of the system ranges from 2μm to 4.32μm. Nevertheless, the lateral resolution of the phase reconstruction can reach up to 1μm for the target with a higher phase shift. It has to be noted that such resolution is achieved with an ideal phase object with homogeneous refractive index distribution.

**3D imaging of biomedical samples through the fiber endoscope**

Both morphological and quantitative biomedical parameters can be extracted from an OPD image, indicating the great potential in biomedical applications of the quantitative phase microendoscope. The cell-like Polyacrylamide (PAAm) hydrogel bead has a spherical shape and homogeneous refractive index distribution [47]. This makes PAAm beads ideal for verifying the OPD measurement fidelity with the microendoscope for biomedical samples.

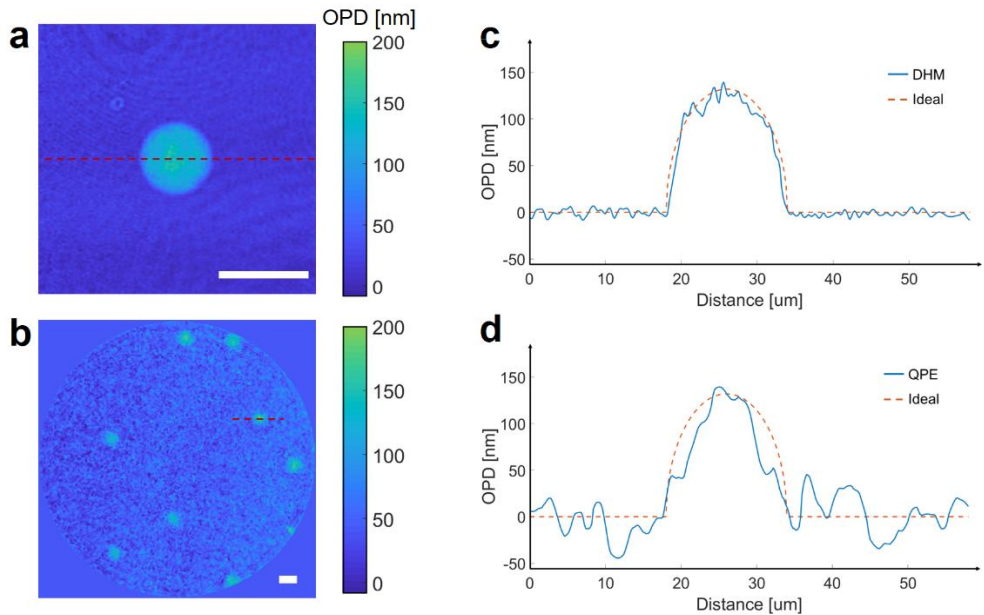

**Fig. 5. Quantitative optical path difference (OPD) map of PAAm micro-gel beads acquired with** (**a**) digital holographic microscope (DHM), (**b**) quantitative phase microendoscope (QPE). Scale bars 20μm. (**c-d**) OPD distribution along the red marked lines in the maps on the left side. The dashed lines represent the ideal OPD distribution of the PAAm bead.

A reference OPD measurement of the bead is done on a digital holographic microscope (DHM) (see Fig. 5a). The average diameter of the beads is measured as 16.7μm. PAAm beads in suspension are also resolved clearly in the reconstructed OPD map with the MCF-based microendoscope (see Fig. 5b). Due to the ideal spherical shape and homogeneous refractive index distribution of the beads, the refractive index of the beads can be calculated precisely from the OPD (Eq. 3). The OPD distribution of the lines (marked in red in Fig. 5a, b)) through the center of the bead are demonstrated as blue lines in Fig. 5c, d. Simulated OPD distribution of an ideal sphere with refractive index difference 0.008 to the background, demonstrated as orange dashed lines in Fig. 5c, d, fits the measured OPD distribution well. The refractive index of the medium (D-PBS) is determined as 1.335, hence the refractive index of the PAAm beads is measured as 1.343. Although the reconstructed OPD map from the microendoscope has a relatively lower spatial resolution and higher background noise, the quantitative OPD values of the beads are recovered correctly. This verifies the strength of the quantitative phase microendoscope to provide accurate OPD measurement at the single-cell level.

Page 9 of 17

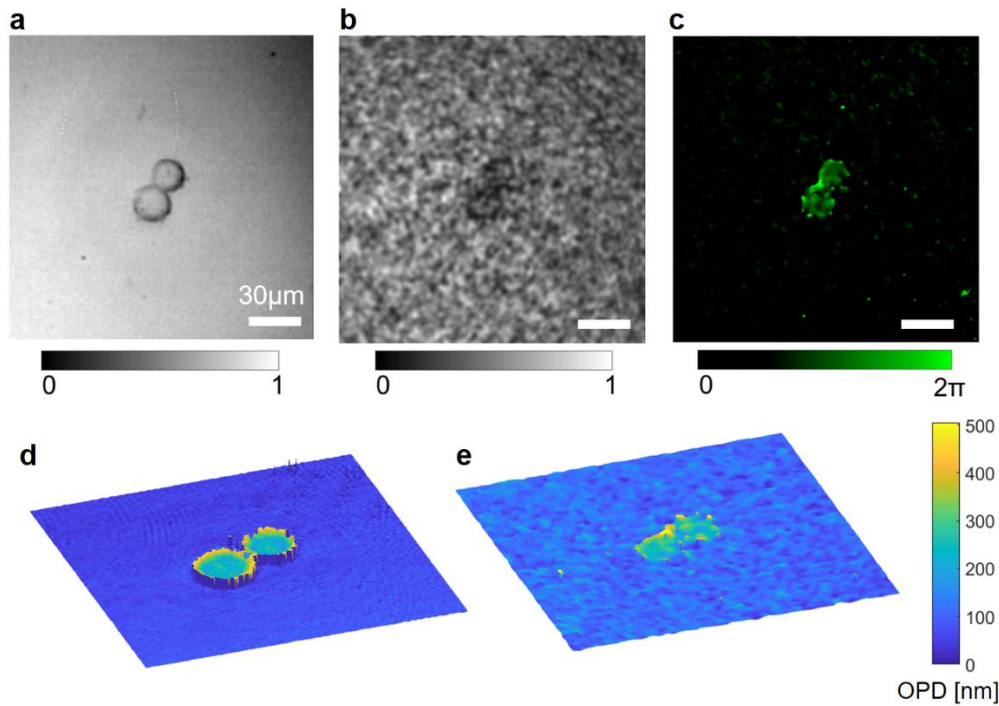

**Fig. 6. Quantitative phase imaging of cancer cell cytokinesis via microendoscope.** (**a**) Intensity image of a HeLa cell in cytokinesis from a conventional reflective microscope. (**b-c**) Microendoscopic (**b**) amplitude and (**c**) phase reconstruction of the HeLa cell in cytokinesis from the far-field speckle. Scale bars 30μm. (**d-e**) 3D OPD map of the HeLa cell measured from the (**d**) conventional digital holographic microscope and the (**e**) quantitative phase microendoscope.

Human cancer cells are used to characterize the performance of the quantitative phase imaging for biological cells via microendoscope. The image of a HeLa cell in cytokinesis captured from a bulky reflective microscope is shown in Fig. 6a. The reconstructed amplitude image of the same cell through the microendoscope is demonstrated in Fig. 6b. The cell is still distinguishable from the background noise. The contrast is significantly improved in the reconstructed phase image (see Fig. 6c). The cancer cell undergoing cytokinesis is clearly resolved by a microendoscope without labeling. It is visible in the phase reconstruction that the cell membranes of two daughter cells are not separated yet. The 3D OPD map of the HeLa cell is thus calculated from the phase shift measured by the quantitative phase microendoscope (see Fig. 6e and Supplementary video V5). A high-resolution OPD map of a similar HeLa cell in cytokinesis is reconstructed from the DHM as a reference measurement (see Fig. 6d and Supplementary video V4).

## Discussion

Our results demonstrate that the ultra-thin lensless MCF endoscope provides high-resolution QPI in hard-to-reach areas. The reconstructed OPD map demonstrated sufficient image quality for morphological evaluations of live human cancer cells. Vital cellular parameters such as cell volume [1], dry mass [13,14], and refractive index [2,48] can also be extracted from the precise OPD map for cytopathology investigation and clinical diagnosis. The precise quantitative phase imaging performance can be hardly achieved using other endoscopes illuminated by incoherent light sources due to the wide spectrum of the light source. Digital holography is a common method for quantitative phase imaging [49,50], however, digital holographic imaging with long MCFs requires complex optical systems and tedious optical alignment [44,51]. Our proposed FAST reconstruction method does not require digital holography, the precise amplitude and phase images of the sample can be recovered from an intensity-only speckle. The 2D image correlation between the amplitude reconstruction with the off-axis holography and the FAST method is 0.998 (see Fig. S6). The quantitative phase of the sample is also precisely retrieved with a lower phase noise level in the background compared to the digital holographic reconstruction (see Fig. S6e, f). Therefore, the holography-free lensless microendoscope based on FAST reconstruction provides sufficient reconstruction quality for both amplitude and phase images with a significantly simplified optical system.

Furthermore, the proposed method enables digital refocusing of the reconstructed complex light field to different depths, which significantly increases the depth of field of the lensless microendoscope to several millimeters and gives more degrees of freedom in sample examinations. The miniature lensless microendoscope with a diameter of 0.35mm is so far the tiniest imaging probe with micrometer range lateral resolution and nanoscale axial sensitivity, paving the way to *in vivo* label-free detection with minimal invasiveness.

Different from our previously reported lensless endoscope based on 3D scanning imaging [41,42], the current setup can achieve comparable lateral resolution and higher axial resolution from the numerical reconstruction of a far-field speckle pattern. It has to be noted that such a scanning-free imaging modality is highly desired for high throughput measurements and dynamic monitoring of samples because the image capturing speed can reach the maximum frame rate of the detection camera. Due to the simplicity of the proposed method, digital holography and wavefront shaping are not necessary, which leads to a compact and cost-effective system. Additionally, compared to endoscopes implementing micro-lens or printed structures on fiber tips [27,29,30], our system is built from a commercially off-the-shelf fiber bundle and common optical components, which is easily replicable for further applications.

The iterative process of the FAST reconstruction may raise concerns about computational time. Due to the complexity of the reference phase shift of the fiber bundle (Fig. 2d), two speckles at different axial distances are required as the input for the FAST reconstruction. The reconstruction is operated on MATLAB and takes about 8 min to calculate the reference phase shift ($2560 \times 1920$ pixels) on a desktop computer (CPU, AMD Ryzen Threadripper 3960X) with GPU (NVIDIA TITAN RTX) acceleration. To reconstruct the further phase shift caused by the sample (Fig. 2d), only a single speckle image is required with the reference phase shift as the initial phase for the iteration, and it only takes about 24 s for a single reconstruction. Performing the reconstruction on multiple GPUs in a parallel pool can further improve the computational time, because the algorithm relies on 2D FFT. The stability of the system when bending the fiber bundle is a critical attribute for *in vivo* applications. Slight deformation of the fiber bundle after the reference measurement would lead to an additional global tilt of the image plane on the detection side [37,52]. The resulting tilt on the phase reconstruction, which is extracted from the background, can be further corrected numerically. Recently reported custom-designed twisted MCF, shows a performance independent of the fiber bending [53], which would further increase the degree of freedom of a lensless microendoscope.

# Materials and Methods

## Multi-core fiber bundle

A 40-cm-long fiber bundle (FIGH-350S, Fujikura, Japan) with around 10,000 cores is used in this work. The diameter of the fiber bundle is 350μm. The average core diameter is 2μm and core-to-core spacing is 3.2μm.

## Experimental setup

The experimental setup is shown in Fig. 1e. A 532nm diode-pumped solid-state continuous-wave laser (Verdi, Coherent Inc., USA) is coupled into a single-mode fiber (460HP, Thorlabs, Germany), and the output beam from the single-mode fiber is used to illuminate the sample. The diffracted light field incident the MCF at the measurement side. On the detection side, a 10× microscope objective (0.25 NA; Plan Achromat Objective, Olympus) and an achromatic lens (f=200mm; Thorlabs, Germany) compose a 4-f system. Hence, the magnified far-field speckle can be projected on the detection camera (UI-3482LE, IDS GmbH, Germany). Due to the random birefringence of the fiber cores, a linear polarizer is placed in front of the camera (LPVISE100-A, Thorlabs, Germany) for capturing a linearly polarized light field.

## Speckle reconstruction algorithm

A specialized phase retrieval algorithm for the fiber bundle is implemented to reconstruct the phase on the fiber facet from the far-field speckle image. The iterative reconstruction process is demonstrated in Fig. S1 and explained in supplementary materials. The total variation minimization algorithm [54] is implemented on the reconstructed amplitude images to reduce the speckle noises. On the other hand, a 2D median filter is applied to the reconstructed phase images to reduce the phase spikes.

## Optical path difference (OPD)

When a coherent light propagates through a homogeneous medium with a refractive index $n$, the OPL is defined as the product of the geometric traveling distance $d$ of light. Therefore, the OPLs are different when the coherent light travels through mediums with different refractive indices $n_0$, $n_1$ at the same distance $d$, and the OPD is defined as

$$\text{OPD} = (n_1 - n_0)d \qquad (3)$$

In experiments, the OPD can be measured from the phase shift $\Delta\phi$ of a coherent light source that passes through mediums with different refractive indices

$$\text{OPD} = (\frac{\Delta\phi}{2\pi} + k)\lambda \qquad (4)$$

where $k$ is non-negative integers, $\lambda$ is the wavelength of the light source. Hence, the phase shift also corresponds to the refractive index difference and the thickness of the medium.

$$\Delta\phi = \frac{2\pi d}{\lambda}(n_1 - n_0) + 2\pi k \qquad (5)$$

**Phase target**

The phase target shown in Fig. 4a is displayed on a spatial light modulator (PLUTO, Holoeye Photonics AG, Germany) and projected on the fiber facet at the measurement side.

**Microgel beads preparation**

The polyacrylamide (PAAm) microgel beads functionalized with fluorescent dye were produced by using a microdroplet generation system and protocol described in a previous study [47]. The continuous phase was a fluorinated oil (HFE-7500, Ionic Liquids Technology, Germany) containing ammonium Kritox® surfactant, N,N,N',N'-tetramethylethylenediamine (TEMED), and acrylic acid N-hydroxysuccinimide ester (Sigma-Aldrich Chemie GmbH, Germany). The dispersed phase was a pre-gel mixture of acrylamide, N,N'-methylenebis acrylamide, ammonium persulphate (Sigma-Aldrich Chemie GmbH, Germany) and Alexa Fluor® 488 Hydrazide (Thermo Fisher Scientific, Germany) dissolved in 10mM Tris-buffer. The flow of the two phases was controlled by a pressure microfluidic controller (Fluigent MFCSTM-EX) and adjusted to obtain beads with a final diameter of about 16μm, analyzed by bright-field microscopy. A ratio of the cross-linking agent to a monomer of 3.25% and a total monomer concentration of 9.9% resulted in beads with Young's modulus of about 6kPa, measured by AFM indentation. The functionalized PAAm beads were washed and re-suspended in 1xPBS and stored at 4ºC until further use. To image the PAAm beads with the microendoscope, the beads are suspended in DPBS (Thermo Fisher, USA) and located 0.5mm away from the measurement fiber facet.

**HeLa cell preparation**

The stable HeLa cell line was kindly provided by the lab of Mariana Medina Sánchez (Leibniz Institute for Solid State and Materials Research). HeLa cells were cultured at 37°C in a humidified atmosphere containing 5% CO2 in Dulbecco's modified Eagle's medium (DMEM) (Thermo Fisher, USA) supplemented with 10% (v/v) fetal bovine serum (FBS) (Thermo Fisher, USA), 100U/mL penicillin, and 100μg/mL streptomycin. HeLa cells were recovered and incubated for 2 weeks before use for spheroids culture. Equal amounts of HeLa cells ($2 \times 10^5$ cells resuspended in 4mL) were added to 3.5cm cell-repellent dishes (Greiner bio-one) after trypsinization and washing with PBS (Thermo Fisher, USA) for preparing spheroids with homogeneous sizes. After two days of maturation, spheroids were separated into different groups and incubated with related treatments. The culture medium was exchanged to DPBS (Thermo Fisher, USA) without phenol red before the measurement.

**Acknowledgments**

We would like to express great appreciation to Dr. Robert Kuschmierz for his contributional discussions and support. We would also like to thank Elias Scharf, Jakob Dremel, David Fernando Ortegón González, and Haoyu Wang for valuable discussions. We thank Dr. Mariana Medina Sánchez for providing the HeLa cells. The support from all the colleagues at MST and BIOLAS is greatly appreciated.

**Funding:**
Deutsche Forschungsgemeinschaft (DFG) grant CZ55/40-1
Tsinghua Scholarship for Overseas Graduate Studies grant 2020023
European Union's Horizon 2020 research and innovation programs No 953121 (project FLAMIN-GO)

**Author contributions:**
Conceptualization: J.S.
Methodology: J.S., J.W.
Investigation: J.S., J.W., N.K., J.C.
Visualization: J.S.
Sample preparation: R.G., S.G., J.S.
Supervision: J.C., J.G., L.C.
Project Management: J.C.
Writing—original draft: J.S.
Writing—review & editing: All authors

**Competing interests:** Authors declare that they have no competing interests.

**Data and materials availability:** The Matlab code of the FAST algorithm is publicly available on Github at https://github.com/Jiawei-sn/FAST.

**Supplementary Materials**

Supplementary Information

Figs. S1 to S6

Supplementary video caption V1 to V6

References (1 to 8)

# Supplementary Information for

## Quantitative phase imaging through an ultra-thin lensless fiber endoscope

Jiawei Sun, Jiachen Wu, Song Wu, Ruchi Goswami, Salvatore Girardo, Liangcai Cao, Jochen Guck, Nektarios Koukourakis, and Juergen W. Czarske

*Corresponding author. Email: jiawei.sun@tu-dresden.de (J.S.);
nektarios.koukourakis@tu-dresden.de (N.K.);
juergen.czarske@tu-dresden.de (J.C.)

**This file includes:**

Supplementary Methods
Figs. S1 to S6
Supplementary video caption V1 to V6
References (1 to 8)

**Other Supplementary Materials for this manuscript include the following:**
Supplementary video V1 to V6

**Supplementary Methods**

**Numerical propagation**

For propagating the light field numerically, both amplitude $A(x, y)$ and phase $\phi(x, y)$ information are required to form the complex light field $E(x, y)$ at the initial axial plane

$$E = A \cdot \exp(i\phi). \tag{1}$$

Based on the angular spectrum method, after a propagation distance of $z$, the complex light field $E_z(x, y)$ can be calculated by the angular spectrum method as

$$E_z = \mathrm{iFFT}\left\{\mathrm{FFT}\{E\} e^{jz\sqrt{(2\pi/\lambda)^2 - (k_x^2 + k_y^2)}}\right\}. \tag{2}$$

where FFT and iFFT represent the fast Fourier transform and the inverse fast Fourier transform, $\lambda$ is the wavelength of the light, $k_x$, $k_y$ are the spatial frequencies. The propagated amplitude $A_z(x, y)$ and phase $\phi_z(x, y)$ can thus be calculated from the complex light field

$$A_z = |E_z|, \quad \phi_z = \arg(E_z). \tag{3}$$

**Far-field amplitude-only speckle transfer (FAST)**

We demonstrate the far-field amplitude-only speckle transfer (FAST) algorithm to reconstruct the random phase distribution on the facet from the far-field speckle images on the detection side. Common phase retrieval algorithm [1–3] is robust for recovering the phase information of simple objects, however, it is difficult to recover the random phase distribution of the 10,000 core of the fiber bundle due to its complexity. The algorithm stagnates near the closest local minimum when we try to recover the phase on the facet. Hence, an algorithm provides faster convergence and stronger constraint is demanded in this case. Inspired by the multi-distance phase retrieval [4], which is initially utilized for increasing the convergence speed of the algorithm, we use two far-field speckle images captured at two axial distances to recover the complex phase distribution on the facet. The principle of the phase retrieval process is demonstrated in Fig. S1.

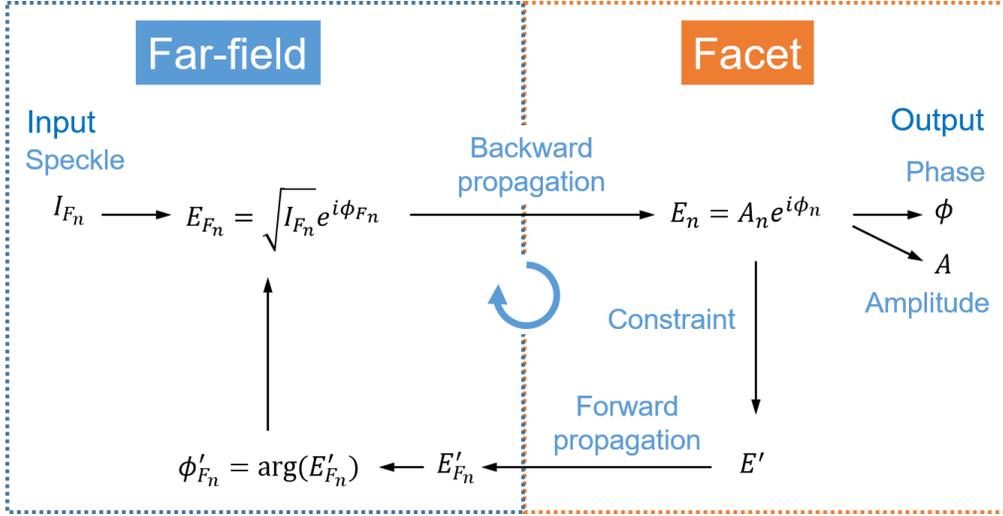

**Fig. S1. Diagram of the iterative phase retrieval algorithm.**

To be more specific, two input speckle intensity images $I_{F_1}(x,y)$, $I_{F_2}(x,y)$ (Fig. S2a) are converted to the amplitude $\sqrt{I_{F_1}(x,y)}$, $\sqrt{I_{F_2}(x,y)}$ of the complex light field in the far field. However, the phase of the complex field is still unknown, hence, random phases $\phi_{F_1}(x,y)$, $\phi_{F_2}(x,y)$ are used to form the estimated far-field complex field $E_{F_1}(x,y)$, $E_{F_2}(x,y)$ at $z_1$, $z_2$.

$$E_{F_n}(x,y) = \sqrt{I_{F_1}(x,y)}\, e^{i\phi_{F_n}(x,y)}. \tag{4}$$

Both complex fields are then back-propagated to the facet plane numerically and two estimated light fields on the facet plane are obtained as

$$E_n = \mathrm{iFFT}\left\{\mathrm{FFT}\left\{E_{F_n}\right\} e^{-jz_n\sqrt{(2\pi/\lambda)^2 - (k_x^2 + k_y^2)}}\right\}. \tag{5}$$

A binary mask $M(x,y)$ (see Fig. S2d), which is obtained by thresholding the captured amplitude image at the facet plane (Fig. S2e), represents the morphology of the fiber bundle facet. The mask is imposed on the estimated light fields on the facet plane to reduce the error in the background. The parameter $\beta$ is used as the weight of the feedback [1], and the estimated complex fields on the facet are modified to

$$E_n^{'} = M \cdot E_n + (1-M)[E^{'}(k-1) - \beta E_n], \tag{6}$$

where $E^{'}(k-1)$ is the mean estimated complex field on the facet from the last iteration, and it is set as a zero-padded image in the first iteration. Therefore, the light fields are updated by the latest field in the masked region and the feedback from the background can be tuned by the weight $\beta$, which is set to 0.2. Then, the mean complex value $E^{'}(x,y)$ of the two estimated fields on the facet plane is calculated by

$$E' = \frac{1}{N}\sum_{n=1}^{N} E'_n, \tag{7}$$

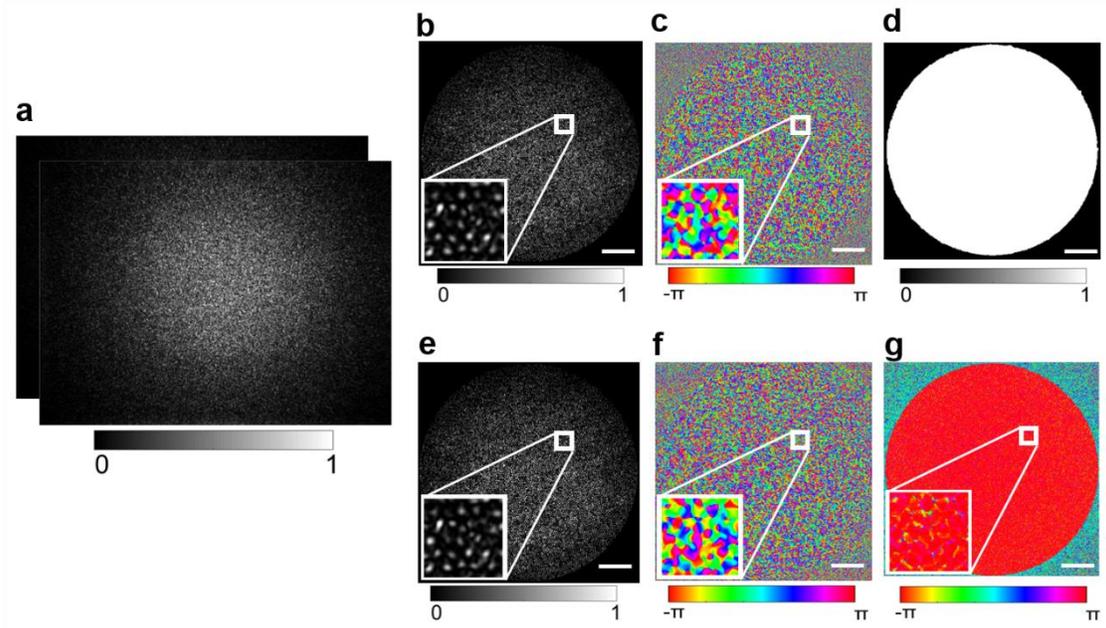

**Fig. S2. Reconstruction of the reference phase distribution on the facet** (**a**) Speckle images at 736μm and 800μm away from the facet. (**b-c**) Reconstructed (**b**) amplitude and (**c**) phase on the facet from the speckles. (**d**) Binary mask used as the weighted constraint. (**e**) Amplitude image of the fiber bundle facet. (**f**) Reconstructed phase on the facet from the off-axis holography. (**g**) Phase difference between (**c**) speckle reconstruction and (**f**) holography reconstruction. Scale bars 50μm.

$N$ denotes the number of speckle images, and $N = 2$ in this case. In the next step, the $E'(x, y)$ is propagated to the far-field located at $z_1$, $z_2$ respectively

$$E'_{F_n} = \text{iFFT}\left\{\text{FFT}\{E'\} e^{jz_n \sqrt{(2\pi/\lambda)^2 - (k_x^2 + k_y^2)}}\right\}. \tag{8}$$

The new estimated phase $\phi'_{F_n}(x, y)$ on both far-field planes can be obtained from the updated complex field $E'_{F_n}(x, y)$

$$\phi'_{F_n}(x, y) = \arg[E'_{F_n}(x, y)]. \tag{9}$$

Hence, the new complex fields in the far-field for the next iteration are updated by the estimated phase, and the original speckle images.

$$E_{F_n}(x, y) = \sqrt{I_{F_n}(x, y)}\, e^{i\phi'_{F_n}(x, y)} \tag{10}$$

The algorithm runs iteratively until the complex light field on the facet is recovered correctly. The correlation coefficient between the reconstructed amplitude on the facet plane $|E'(x, y)|$ and the captured facet image is used as the figure of merit to characterize the performance of the algorithm. The reconstructed amplitude image on the facet after 2,200 iterations is demonstrated in Fig. S2b, which has a correlation coefficient of 0.96 compared with the amplitude image of the facet captured by the microscope (Fig. S2e). To characterize the fidelity of the phase reconstruction, the ground truth phase distribution on the facet (Fig. S2f) is measured with off-axis holography[5] and the circular standard deviation [6] of the difference between the reconstructed phase and the ground truth in the facet region is used as the loss function. The phase deviation also converges when the amplitude correlation converges, and the tendencies of both criteria are very similar in the iteration process. Therefore, it is sufficient to use the amplitude correlation as the exit criteria to get an accurate phase reconstruction. The final reconstructed phase distribution on the facet is demonstrated in Fig. S2c, and the circular standard deviation of the phase difference to the ground truth is 0.41 rad in the facet region, which is 6.5% of $2\pi$. Although there is still noticeable phase noise in the facet area, those noise mostly appears in the area of claddings (see Fig. S2g) which are not used to guide the light.

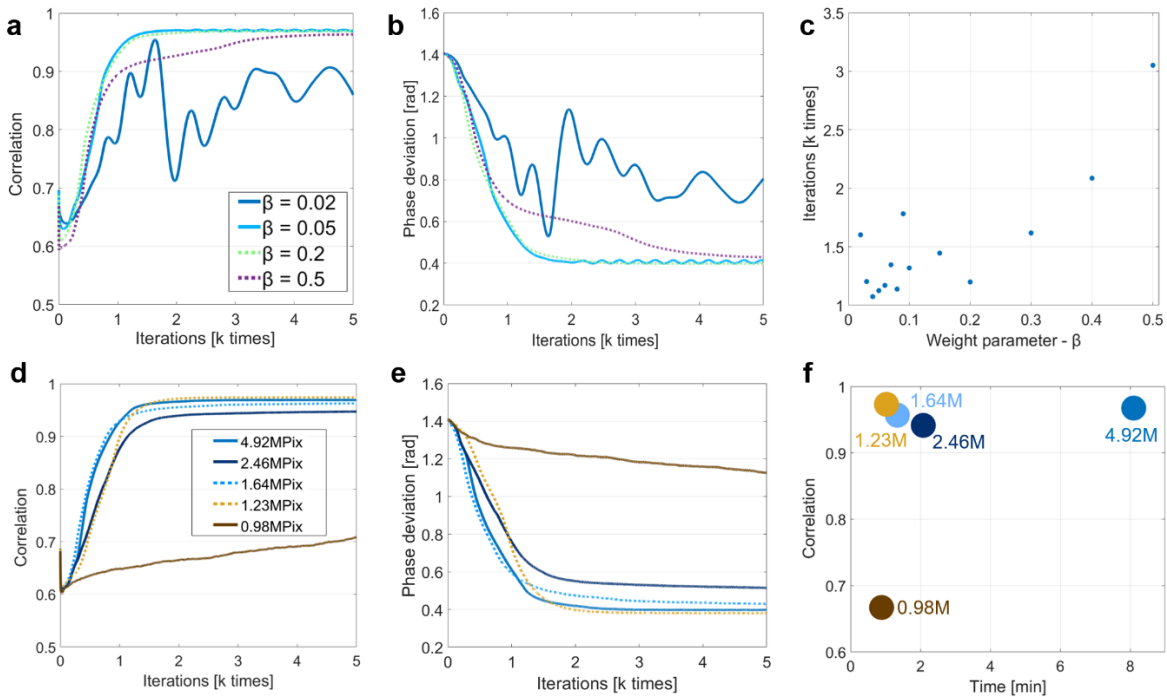

**Fig. S3.** (**a-c**) Comparison of the performance of the algorithm with different weight $\beta$. (**a**) The correlation coefficient between the reconstructed amplitude image and the captured image on the facet is used as the merit function. (**b**) The circular standard deviation of the phase difference between the reconstructed and holographic measured phase is used as the loss function. (**c**) The number of iterations to convergence with different weight parameter $\beta$. (**d-f**) Comparison of the performance of the algorithm with full camera pixels and 1/2, 1/3, 1/4, 1/5 sampled pixels. (**d**) The correlation coefficient is used as the merit function. (**e**) The circular standard deviation of the phase

difference is used as the loss function. **(f)** Computation time and the correlation of the reconstruction for 2,200 iterations.

To characterize the performance of the reconstruction algorithm, the reconstruction process is performed and analyzed with different parameters. In Fig. S3, the correlation between the reconstructed amplitude and the captured amplitude image of the facet represents the fidelity of the reconstruction while the circular standard deviation of the phase difference between the reconstruction and holography measurement represents the error of the reconstruction. The amplitude correlation curve is highly correlated to phase deviation curves in different cases. The weight parameter $\beta$ in Eq. 6 is used to tune the feedback of the estimated light field in the background region. As shown in Fig. S3a and b, the algorithm can hardly converge and oscillates randomly when $\beta \leq 0.02$. For $0.03 \leq \beta \leq 0.06$, the algorithm can converge to a small range but there are still slight oscillations after the convergence. The algorithm converges smoothly for $\beta \geq 0.07$. The required iteration time to achieve an amplitude correlation of 0.95 for different weight $\beta$ is demonstrated in Fig. S3c. It can be noticed that the convergence speed decreases when $\beta \geq 0.3$, therefore, an optimal $\beta = 0.2$ is chosen to achieve smooth and fast convergence.

The camera pixel size and the total number of pixels are both critical to achieving sufficient spatial sampling of the light field. According to the Nyquist–Shannon sampling theorem, to avoid aliasing and resolve the fiber cores with a diameter of $d_{core}$, the camera pixel size $a_{cam}$ should at least fulfill

$$a_{cam} < \frac{m \cdot d_{core}}{2}, \qquad (11)$$

where $m$ is the magnification ratio of the imaging system. In this work, the fiber core diameter is around 2μm and the magnification ratio of the imaging system is 10×. Therefore, the pixel size of the detection camera should be smaller than 10μm. The camera used in this work (UI-3482LE, IDS GmbH, Germany) has 4.92 million (2560×1920) pixels and the pixel size is 2.2μm.

To evaluate the influence of the quantity and size of camera pixels on the reconstruction performance. The speckle images are down-sampled to 2.46 million (1280×960), 1.64 million (853×640), 1.23 million (640×480), and 0.98 million (512×384) pixels. The pixel size set in the algorithm is also increased 2×, 3×, 4×, and 5× to keep the physical size of the image field the same. As shown in Fig. S3d and e, the performance of the algorithm is similar when the camera pixel size fulfills Eq. 11. When the speckle images are sampled to 512×384 pixels and the pixel size is increased to 11μm, the reconstruction can hardly converge. The computation time for 2,200 iterations and the correlation coefficient between the reconstructed amplitude image and the sampled ground truth image with the same pixel number is demonstrated in Fig. S3f. The computation time is decreased significantly with fewer pixels and the reconstruction fidelity remains high if Eq. 11 is fulfilled. It has to be noted that this result only denotes the performance of the reconstruction algorithm is robust when the camera sampling frequency is above the Nyquist frequency, and more camera pixels lead to higher spatial resolution.

**Spatial resolution measurement**

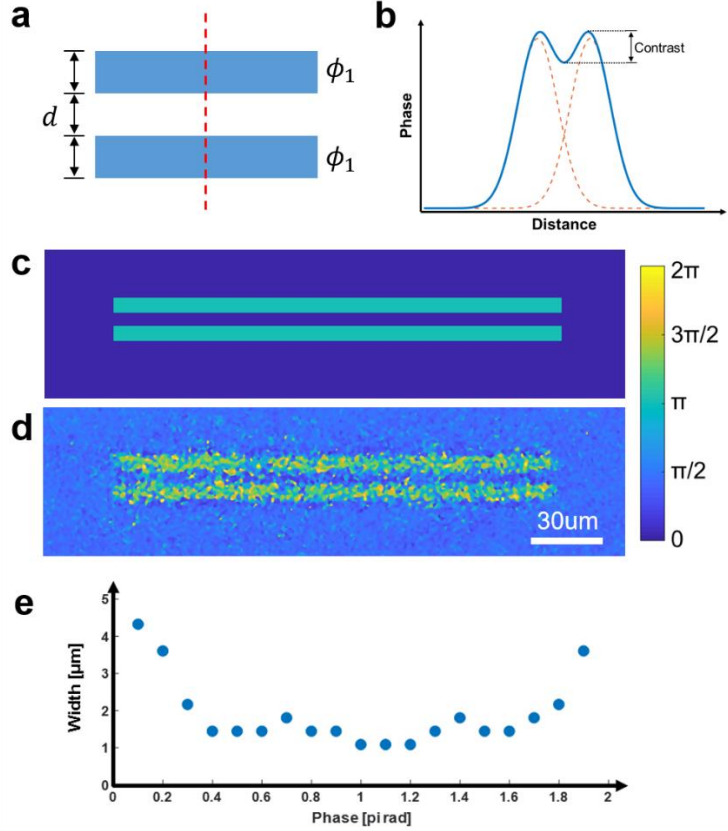

**Fig. S4. The principle of quantifying the spatial resolution of the phase microendoscope.** (**a**) A stripe pattern is used for the resolution test. The upper and bottom band shares the same width and phase, and the gap between the two bands has the same width of the lines. The red dashed line indicates the sampling direction of the phase profile. (**b**) The reconstructed phase profile in the marked direction. (**c**) The phase target displayed on the SLM. (**d**) Reconstructed phase image from the microendoscope. Scale bar 30μm. (**e**) The measured lateral resolution of the system for different phase values.

A program-controlled phase target shown in Fig. S4a is used to characterize the spatial resolution of the quantitative phase microendoscope. The length of the gap between the upper and bottom band is the same as the width of each band. An example of the phase distribution in the marked direction is shown in Fig. S4b. The contrast $\phi_{contrast}$ of the reconstructed phase profile is defined as the phase difference between the maximum and minimum phase value in the gap between the two bands. We define the contrast ratio (CR) as

$$CR = \frac{\phi_{contrast}}{\phi_{max}}, \tag{12}$$

where the $\phi_{max}$ is the maximum phase value. In ideal circumstances, the band is defined as resolvable when the CR is above the Rayleigh criterion, which is 27% here[7].

To evaluate the resolution limit of our system at different phase values, this program-controlled phase target is displayed on a phase-only spatial light modulator (SLM; Pluto, Holoeye GmbH, Germany). As shown in Fig. S4c, the phase target is projected on the imaging plane of the MCF by a telescope system and the pixel size of the SLM is minified to 360nm. In each measurement process, the phase difference between the band and the background is set to a certain value. Firstly, the width of the band is set to one pixel, which corresponds to 360nm, and the corresponding speckle image is recorded on the detection side of the MCF. The phase image is thus reconstructed by the FAST method, and an example of the reconstructed phase image is shown in Fig. S4d. Hence, the CR of the reconstructed phase target is extracted from the corresponding phase profile sampled from the phase reconstruction. The width of the band is further increased in a step of one pixel until the CR of the reconstructed phase target is above the Rayleigh criterion. For instance, when the phase value of the band is set to π and the width of the band is two pixels (0.72μm), the CR of the reconstructed phase target is 24.2%. Then the width is increased to three pixels (1.08μm), and the CR is increased to 51.4%, which means the resolution limit of the system is 1.08μm for an ideal homogeneous object with a phase shift of π. We measured the resolution limit for phase value from 0.1π to 1.9π at a step of 0.1π, and the result is demonstrated in Fig. S4e and Fig. 4f.

**Quantitative phase measurement on a digital holographic microscope (DHM)**

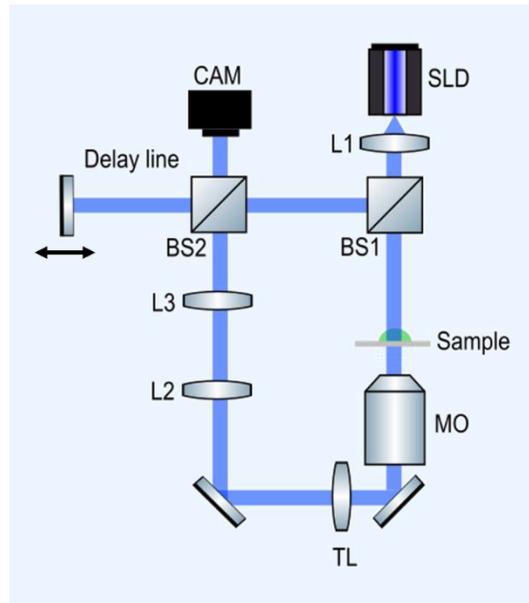

**Fig. S5. Experimental setup of the digital holographic microscope.** SLD, superluminescent diode; L1-L3, lenses; TL, tube lens; BS1, BS2, beamsplitters; MO, microscope objective; CAM, camera.

A DHM is implemented as a reference tool to measure the quantitative phase shift of the sample. The optical setup of the DHM is shown in Fig.S5. A superluminescent diode (SLD; 450nm; EXALOS AG, Switzerland), a low-coherence light source, is implemented to reduce the speckles in the phase reconstructions. A 40x microscope objective (0.65 NA; Plan Achromat Objective,

Olympus) is used to image the PAAm beads shown in Fig. 5a in the main text. A 20× microscope objective (0.4 NA; Plan Achromat Objective, Olympus) is used to image the HeLa cells shown in Fig. 6d in the main text. The DHM is based on a Michelson interferometer. To match the optical path length of the object and reference beam, a mirror is mounted on a motorized stage for dynamic precise control of the path length of the reference beam. The second beamsplitter (BS2) is slightly tilted for off-axis holography, and the holographic reconstruction is based on the spatial filtering[5] and angular spectrum method[8]. The reconstructed phase image is subtracted by the background to get the quantitative phase shift. The corresponding 3D OPD maps in Fig. 5a and 6d are calculated from the phase shift.

**Two-dimensional (2D) image correlation coefficient**
The 2D correlation coefficient is employed to characterize the fidelity of reconstructed amplitude images. Hence, for a normalized image $X(x, y)$, the correlation coefficient (CC) between the reference image $Y(x, y)$ is expressed as

$$\text{CC} = \frac{\sum_{i}^{n}(X_i - \bar{X})(Y_i - \bar{Y})}{\left\{\sum_{i}^{n}(X_i - \bar{X})^2 \sum_{i}^{n}(Y_i - \bar{Y})^2\right\}^{1/2}}, \qquad (13)$$

where $\bar{X}(x, y)$ and $\bar{Y}(x, y)$ is the mean value of the reconstructed image and the reference image respectively, and $n$ is the total number of pixels.

**Glass beads flow**
1mg glass beads (Dantec Dynamics, Denmark) are suspended in 0.5ml Dulbecco's phosphate-buffered saline (DPBS) (Thermo Fisher, USA). Sizes of glass beads vary from 2μm to 20μm and the mean diameter is 10μm. The suspension is pumped into a microchannel constantly by a syringe. The height of the microchannel (μ-Slide Luer, ibidi GmbH, Germany) is 100μm and the channel volume is 25μl.

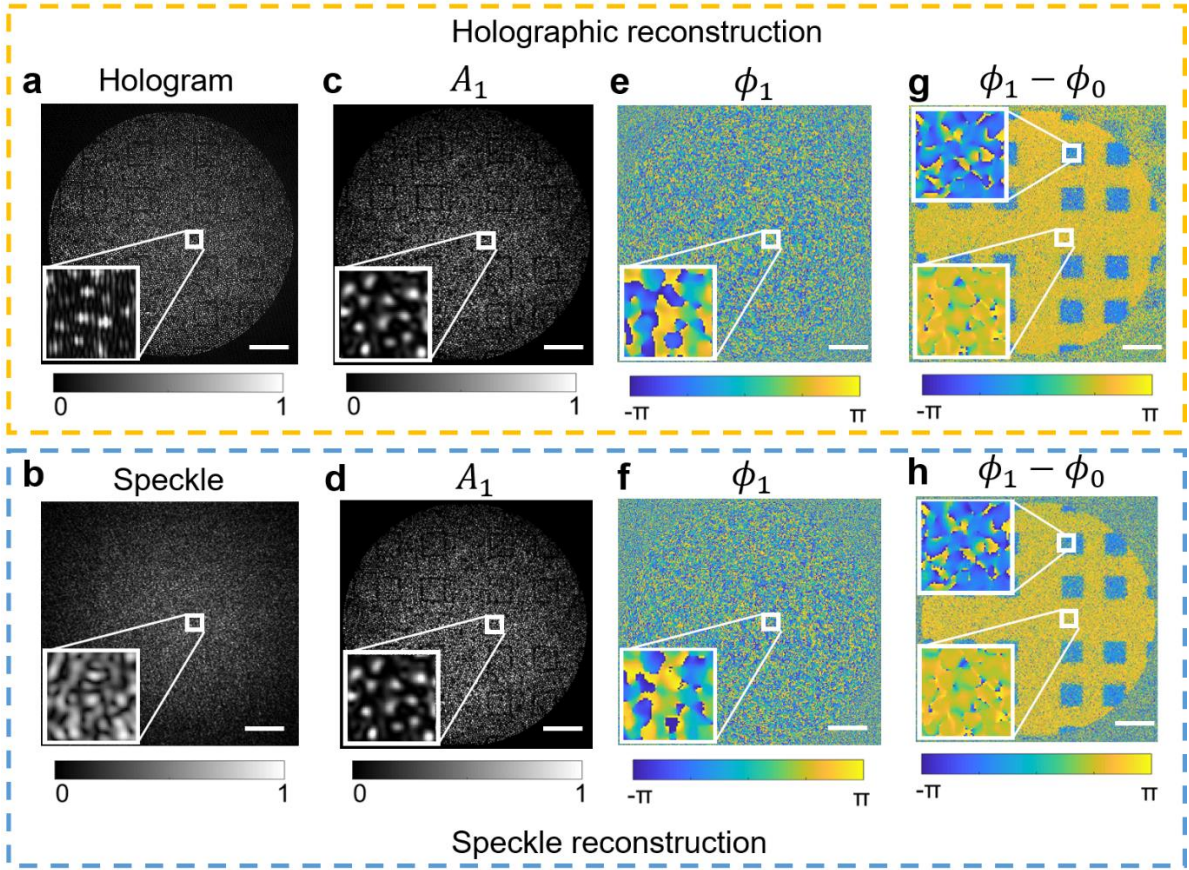

**Fig. S6. Comparison of off-axis holographic reconstruction and far-field speckle reconstruction.** (**a**) Off-axis hologram of a phase target captured at the facet on the detection side. (**b**) Far-field speckle image of the phase target captured on the far-field of the detection side. (**c-d**) Reconstructed amplitude image from the (**c**) hologram (**d**) speckle. (**e-f**) Reconstructed phase image from the (**e**) hologram (**f**) speckle. (**g-h**) Quantitative phase images of the target are retrieved from (**g**) holographic (**h**) speckle reconstruction. Scale bars 50μm.

**Supplementary video V1.**

Digital focusing of the multi-layer target in axial distances from 0 to 5mm away from the fiber facet at the measurement side. The top layer is digitally refocused at 1.26mm and the bottom layer is focused at 2.66mm away from the fiber facet.

**Supplementary video V2.**

Glass beads flowing through a microchannel are imaged by the lensless microendoscope at the measurement side. The speckle images, which are the system response shown on the left side of the video, are recorded by the detection camera at a video rate. The reconstructed video of the glass bead flow is shown on the right side of the video.

**Supplementary video V3.**

Reconstructed 3D optical path difference map of the phase target imaged through the quantitative phase microendoscope.

**Supplementary video V4.**

Reconstructed 3D optical path difference map of the HeLa cell in cytokinesis via the digital holographic microscope.

**Supplementary video V5.**

Reconstructed 3D optical path difference map of the HeLa cell in cytokinesis imaged through the quantitative phase microendoscope.

Supplementary information accompanies the manuscript on the Light: Science & Applications website (http://www.nature.com/lsa).